\documentclass[prd,nofootinbib,floats,superscriptaddress,eqsecnum,tightenlines,11pt]{revtex4}

\usepackage{hyperref}
\usepackage{graphicx}
\usepackage{amsmath,amssymb,amsfonts,amsthm,latexsym,stmaryrd}
\usepackage{marginnote}
\usepackage{color}

\def\beq{\begin{equation}}
\def\eeq{\end{equation}}
\newcommand{\bea}{\begin{eqnarray}}
\newcommand{\eea}{\end{eqnarray}}
\def\bi{\begin{itemize}}
\def\ei{\end{itemize}}
\def\ba{\begin{array}}
\def\ea{\end{array}}
\def\bfig{\begin{figure}}
\def\efig{\end{figure}}
\def\d{\delta}

\def\tA{\hat A} 

\def\a{\alpha}
\def\f{f}

\def\aa{\alpha_1}
\def\ab{\alpha_2}
\def\ac{\alpha_3}
\def\ad{\alpha_4}
\def\ae{\alpha_5}
\def\ba{\beta_1}
\def\bb{\beta_2}

\def\ka{\kappa_1}
\def\kb{\kappa_2}
\def\kc{\kappa_3}
\def\kd{\kappa_4}
\def\ke{\kappa_5}

\def\An{A_*}

\def\A{{\cal A}}

\def\R{R}
\def\tR{\tilde R}
\def\h{h}

\def\C{A} 
\def\D{B} 

\def\tf{\tilde f}
\def\tg{\tilde g}
\def\tnabla{\tilde \nabla}
\def\T{{\cal T}}
\def\tf{\tilde f}
\def\ta{\tilde\alpha}
\def\r{{\gamma}}
\def\tX{\tilde X}
\def\l{\lambda}

\def\fX{f_{,X}}

\newcommand{\Gfour}{G_4{}}
\newcommand{\Ffour}{F_4{}}

\begin{document}

\title{Degenerate higher order scalar-tensor theories beyond Horndeski and disformal transformations}
\author{Jibril Ben Achour}
\email{jibrilbenachour@gmail.com}
\affiliation{Laboratoire APC -- Astroparticule et Cosmologie, Universit\'e Paris Diderot Paris 7, 75013 Paris, France}
\affiliation{Department of Physics \& Center for Field Theory and Particle Physics, Fudan University, 200433 Shanghai, China}
\author{David  Langlois}
\email{langlois@apc.univ-paris7.fr}
\affiliation{Laboratoire APC -- Astroparticule et Cosmologie, Universit\'e Paris Diderot Paris 7, 75013 Paris, France}
\author{Karim Noui}
\email{karim.noui@lmpt.univ-tours.fr}
\affiliation{Laboratoire de Math\'ematiques et Physique Th\'eorique, Universit\'e Fran\c cois Rabelais, Parc de Grandmont, 37200 Tours, France}
\affiliation{Laboratoire APC -- Astroparticule et Cosmologie, Universit\'e Paris Diderot Paris 7, 75013 Paris, France}

\date{\today}

\begin{abstract}
We consider all degenerate scalar-tensor theories that depend quadratically on second order derivatives of a scalar field, which we have identified in a previous work. These theories,  whose degeneracy in general
  ensures the absence of Ostrogradski instability, include the quartic Horndenski Lagrangian as well as its quartic  extension beyond Horndeski, but also other families of Lagrangians. We study how all these theories transform under general conformal-disformal transformations  and find that they can be separated into three main classes that are  stable under these transformations. This leads to a  complete classification modulo conformal-disformal transformations.  Finally, we show that these higher order theories include mimetic gravity and  some particular khronometric theories. They also contain  theories that  do not correspond, to our knowledge,  to already studied theories,  even up to field redefinition.
\end{abstract}

\maketitle

\section{Introduction}

Scalar tensor theories play a prominent role in theories of modified gravity. As ever more sophisticated  models have been considered, special attention was lately devoted to scalar tensor Lagrangians that contain second order derivatives of a scalar field. A crucial requirement for such theories is the absence of the so-called Ostrogradski ghost, in order to avoid disastrous instabilities~\cite{Ostrogradski}. It has been thought for a long time that the absence of an Ostrogradski ghost demands the Euler-Lagrange equations  to be at most second-order, which explains why the literature has been mostly limited to the study of Horndeski's theories~\cite{horndeski,Deffayet:2011gz} until recently. But the discovery of  viable theories ``beyond Horndeski''~\cite{Zumalacarregui:2013pma,Gleyzes:2014dya,Gleyzes:2014qga}, i.e. possessing Euler-Lagrange equations for the metric and scalar field whose order is higher than two, has challenged this preconception. 

A higher order scalar tensor theory generically contains four degrees of freedom,  including the  Ostrogradski ghost. As we proposed in \cite{LN1}, a systematic way to identify scalar-tensor theories  that contain only  three degrees of freedom is to consider  Lagrangians that are degenerate, in a generalized sense involving the coupling between the metric and the scalar field. From the Hamiltonian point of view, this degeneracy implies the existence of  phase space constraints, in addition to the usual Hamiltonian and momentum constraints due to diffeomorphism invariance, and explains why one degree of freedom is eliminated, even  if the equations of motion are higher order. A detailed Hamiltonian analysis has confirmed the direct link between this degeneracy and the elimination of the  Ostrogradski ghost~\cite{LN2}. 

The degeneracy criterium, which provides  a powerful and simple method to identify viable theories,  was used  in  \cite{LN1} to  find all scalar tensor theories based on a Lagrangian quadratic in second order derivatives of a scalar field,  together with a term proportional to the scalar curvature.  
Within these degenerate higher order  scalar tensor (DHOST) theories, we recovered, as particular cases, the (quadratic) Horndeski Lagrangian $L_4^{\rm H}$ as well as its extension $L_4^{\rm bH}$ introduced in \cite{Gleyzes:2014dya,Gleyzes:2014qga} (see section \ref{sec_STT} for an explicit definition of these Lagrangians). We also considered the quintic extension beyond Horndeski of \cite{Gleyzes:2014dya}, $L_5^{\rm bH}$ , which is degenerate by itself or combined with $L_4^{\rm bH}$ but not with an arbitrary $L_4^{\rm H}$. By using  the same degeneracy argument, the combinations involving $L_5^{\rm H}$ too were studied in \cite{Crisostomi:2016tcp}. 
In particular, the results of \cite{LN1, Crisostomi:2016tcp} show  that only specific combinations of Horndeski Lagrangians with their quartic and quintic extensions beyond Horndeski are viable and they  coincide with the combinations obtained in \cite{Gleyzes:2014qga} via disformal transformation of Horndeski.

The goal of the present work is to examine in more detail all the quadratic  DHOST theories of \cite{LN1} and investigate whether they can be related, or not,  to already known theories  via  generalized disformal transformations~\cite{Bekenstein:1992pj}, i.e. redefinitions of the metric of the form 
\beq
\label{g_disf}
\tg_{\mu\nu}=\C(X,\phi) g_{\mu\nu} +\D(X,\phi) \nabla_\mu\phi \nabla_\nu\phi\,,
\eeq
where $X\equiv g^{\mu\nu}\nabla_\mu\phi\nabla_\nu\phi$.
Several results concerning the disformal transformations of Horndeski theories have already been established in previous works. It was shown, in  \cite{Bettoni:2013diz}, that Horndeski theories transform into themselves under special disformal transformations where $\C$ and $\D$  depend  on $\phi$ only,  not on $X$. 
The general disformal transformation of the Einstein-Hilbert Lagrangian was computed  in \cite{Zumalacarregui:2013pma}, providing the first example of theory ``beyond Horndeski'', i.e. a ghost-free theory with higher order Euler-Lagrange equations of motion. In \cite{Gleyzes:2014qga}, it was shown that disformal transformations of Horndeski theories with $\C=1$ lead to the extensions beyond Horndeski proposed in \cite{Gleyzes:2014dya}. The $X$-dependent disformal transformations have  also been studied recently in several papers (see e.g. \cite{Motohashi:2015pra,Domenech:2015hka,Tsujikawa:2015upa,Domenech:2015tca,Emond:2015efw}, and  \cite{Fujita:2015ymn} for scalar tensor theories that explicitly  break  spacetime covariance~\cite{Gao:2014soa}.)

In this work, we present  the general disformal transformation of all quadratic DHOST theories identified in \cite{LN1}. This is useful to show that the  three main classes of theories, as well as the few subclasses within each,  are stable under disformal transformations. 
Part of our results coincides with the conclusions of \cite{Crisostomi:2016czh}, which also studies the theories of \cite{LN1} and  especially the class of theories  related to (quadratic) Horndeski via disformal transformations. Here, we derive  the  transformation laws of the arbitrary functions in the general action, which  enables us to study the  disformal transformations in the other classes as well. 

Interestingly,  the quadratic DHOST theories contain a few theories which have been well studied in the literature. Indeed, the Lagrangians that remain invariant under a field redefinition of the scalar field correspond to khronometric theories~\cite{Blas:2010hb}, which are a subset of Einstein-Aether theories~\cite{Jacobson:2000xp}. Note that khronometric theories are not in general degenerate  and only a subset of them appear among DHOST theories. Finally, we also discuss mimetic gravity~\cite{Chamseddine:2013kea} and related theories~\cite{Deruelle:2014zza}, which are obtained from the Einstein-Hilbert action by a disformal transformation that is not invertible (see also \cite{Zumalacarregui:2013pma}).

Our paper is organized as follows. In the next section, we introduce the general form  of the Lagrangians we will study. In section \ref{section_degenerate}, we summarize  the main results obtained in \cite{LN1} and present the classification of quadratic DHOST theories.   In section \ref{section_disformal}, we derive the general disformal transformation of any  quadratic DHOST Lagrangian. This enables us to show that all  classes  are stable under these transformations. In section \ref{section_classI}, we consider the theories related to Horndeski. Other classes are analysed in the subsequent section. In section \ref{section_khronon_mimetic}, we discuss the degenerate khronometric theories as well as mimetic theories. We conclude in the final section. 
 
\section{Scalar tensor theories}
\label{sec_STT}
\subsection{The action}
In this work, we  consider scalar-tensor theories  whose dynamics is governed by an action of the general form
\beq
\label{action}
S=S_{g}+S_\phi,
\eeq
where the first contribution involves the Ricci scalar $\R$ of the metric $g_{\mu\nu}$,
\beq
\label{S_g}
S_g\equiv \int d^4x\,  \sqrt{-g}\,  \f(\phi, X) \, \R \,,
\eeq
and the second contribution  depends quadratically on the second derivatives of the scalar field $\phi$
\beq
\label{S_phi0}
S_\phi \equiv \int  d^4x\,\sqrt{- g}\,   C^{\mu\nu,\rho\sigma}\,  \nabla_\mu\! \nabla_\nu\phi \  \nabla_\rho \!\nabla_\sigma\phi\,,
\eeq
$C^{\mu\nu,\rho\sigma}$ being an arbitrary tensor that depends only on $\phi$ and $\nabla_\mu\phi$. Note that $S_{g}$ reduces to the familiar Einstein-Hilbert action when the function $\f$ is  constant.

We stress that our analysis is also valid if we add to the above action extra contributions that depend at most linearly on $\phi_{\mu\nu}$, i.e. of the form
\beq
\label{S_other}
S_{\rm other}=\int  d^4x\,\sqrt{-g}\,  \left\{P(\phi, X) + Q_1(\phi, X) g^{\mu\nu}\phi_{\mu\nu}
+Q_2(\phi, X) 
\, \phi^\mu \phi_{\mu\nu}\phi^\nu \right
\}\,,
\eeq
where we have used the compact notation $\phi_\mu\equiv \nabla_\mu\phi$ and $\phi_{\mu\nu}\equiv \nabla_\mu\! \nabla_\nu\phi$. 
  These additional contributions do not  modify the degeneracy conditions derived in \cite{LN1}, which will be summarized in the next section. For simplicity, we will not include these terms explicitly in our study but one should keep in mind that they can be present.

Without loss of generality, we require the tensor $C^{\mu\nu,\rho\sigma}$ in (\ref{S_phi0}) to satisfy the index symmetries 
\beq
C^{\mu\nu,\rho\sigma} = C^{\nu\mu,\rho\sigma}= C^{\mu\nu,\sigma\rho}= C^{\rho\sigma,\mu\nu}\,,
\eeq 
which implies that  the most general form of this tensor  is
\begin{eqnarray}\label{family}
C^{\mu\nu,\rho\sigma} & = &  \frac{1}{2} \aa\, (g^{\mu\rho} g^{\nu\sigma} + g^{\mu\sigma} g^{\nu\rho})+
\ab \,g^{\mu\nu} g^{\rho\sigma} +\frac{1}{2} \ac\, (\phi^\mu\phi^\nu g^{\rho\sigma} +\phi^\rho\phi^\sigma g^{\mu\nu} ) 
\cr
& & \quad +   \frac{1}{4} \ad (\phi^\mu \phi^\rho g^{\nu\sigma} + \phi^\nu \phi^\rho g^{\mu\sigma} + \phi^\mu \phi^\sigma g^{\nu\rho} + \phi^\nu \phi^\sigma g^{\mu\rho} ) +  \ae\, \phi^\mu \phi^\nu \phi^\rho \phi^\sigma \label{four} \,,
\end{eqnarray}
where the $\a_I$ are five arbitrary functions of $\phi$ and $X$. Defining the five  elementary Lagrangians quadratic in second derivatives
\begin{eqnarray}
L_1^\phi&\equiv& \phi^{\mu\nu} \phi_{\mu\nu}\,, \quad 
L_2^\phi\equiv (\phi_\mu^{\ \mu})^2\,, \quad 
L_3^\phi\equiv  \phi_\mu^{\ \mu} \, \phi^\rho\phi_{\rho\sigma}\phi^\sigma\,,
\nonumber
\\
L_4^\phi &\equiv& \phi^\mu \phi_{\mu\nu}\phi^{\nu\rho}\phi_\rho
\,, \quad 
L_5^\phi\equiv (\phi^\rho\phi_{\rho\sigma}\phi^\sigma)^2
\,, 
\end{eqnarray}
 the action $S_\phi$  in (\ref{S_phi0}) now reads
\beq
\label{S_phi}
S_\phi = \int d^4x \sqrt{- g}\, \left(\aa L_1^\phi +\ab L_2^\phi+\ac L_3^\phi+\ad L_4^\phi
+\ae L_5^\phi
\right)\equiv  \int d^4x \sqrt{- g}\, \a_I L_I^\phi\,,
\eeq
where the summation over the index $I$ ($I=1,\dots, 5$) is implicit in the last expression.

\subsection{Other curvature terms}
It is not difficult to see  that the general action (\ref{action}) also includes terms of the form
\beq
\label{S_Ricci}
S_{\rm Ricci}\equiv \int d^4x \sqrt{-g} \, h(\phi, X) \, \R_{\mu\nu}\phi^\mu \phi^\nu\,,
\eeq
where $h$ is an arbitrary function.
Indeed, using the definition of the Ricci tensor and the properties of the Riemann tensor, one can write
\begin{eqnarray}
\phi^\mu \R_{\mu\nu} \phi^\nu&=& -\phi^\mu g^{\rho\sigma}\R_{\rho\mu\nu\sigma} \phi^\nu=
-\phi^\mu g^{\rho\sigma}\left(\nabla_\rho\nabla_\mu-\nabla_\mu\nabla_\rho\right)\phi_\sigma
\cr
&=&
-\phi^\mu\nabla_\mu \nabla_\nu \phi^\nu+\phi^\mu\nabla_\nu \nabla_\mu \phi^\nu\,.
\end{eqnarray}
Substituting this into the action (\ref{S_Ricci}), one gets, after integration by parts,
\beq
S_{\rm Ricci}\equiv \int d^4x \sqrt{-g}\left\{ \, -h \left(L_1^\phi-L_2^\phi\right) +2 h_X\left(L_3^\phi-L_4^\phi\right)+h_\phi\left(X\phi^\mu_\mu -\phi^\mu \phi_{\mu\nu}\phi^\nu\right)
\right\}\,,
\eeq
where the contribution proportional to $h_\phi$ is of the form (\ref{S_other}).

\subsection{Particular cases}
The  theories (\ref{action}) include as a particular case the quartic Horndeski term
\beq
L^{\rm H}_4 = \Gfour(\phi,X) \, \R - 2 \Gfour_{,X}(\phi,X) (\Box \phi^2 - \phi^{ \mu \nu} \phi_{ \mu \nu}) \,,
\eeq
which corresponds to  (\ref{S_g}) and (\ref{S_phi}) with 
\beq
f=\Gfour\,, \qquad \aa= -\ab= 2 \Gfour_{,X}\,, \qquad \ac=\ad=\ae=0\,.
\eeq

The action (\ref{action}) also includes the extension beyond Horndeski introduced in \cite{Gleyzes:2014dya}, which can be written as
\beq
L^{\rm bH}_4=\Ffour(\phi,X) \epsilon^{\mu\nu\rho}_{\ \ \ \ \sigma}\, \epsilon^{\mu'\nu'\rho'\sigma}\phi_{\mu}\phi_{\mu'}\phi_{\nu\nu'}\phi_{\rho\rho'}\,.
\eeq
This is of the form (\ref{S_phi}) with 
\beq
\aa=-\ab= X \Ffour     \,, \qquad \ac=-\ad= 2 \Ffour\,, \qquad \ae=0\,.
\eeq
Of course, any combination of $L^{\rm H}_4 $ and $L^{\rm bH}_4$ is also among the theories (\ref{action}).

 \section{Classification of degenerate theories}
 \label{section_degenerate}
 In this section, we  summarize  the main results obtained in  \cite{LN1}, as well as some additional elements derived in \cite{LN2}, and present all  the quadratic DHOST theories, i.e. all the theories of the form (\ref{action}) which are degenerate.
 
 \subsection{Degeneracy conditions}
 In order to study the degeneracy of (\ref{action}), it is useful to introduce the auxiliary field $A_\mu\equiv \nabla_\mu\phi$. For an arbitrary foliation of spacetime by spacelike hypersurfaces $\Sigma(t)$, endowed with spatial metric $\h_{ij}$, the metric in ADM form reads 
 \beq
 ds^2=-N^2 dt^2 +\h_{ij} (dx^i+N^i dt)(dx^j+N^jdt)\,,
 \eeq
 where $N$ is the lapse and $N^i$ the shift vector. 
 The $(3+1)$ decomposition of the action (\ref{action}) leads to a kinetic term of the form~\cite{LN1}
 \beq
 \label{S_kin}
S_{\rm kin}   =  \int  dt \, d^3x \, N \sqrt{\h} \left[\frac{1}{N^2}{\cal A} \, A_*^2 +
\frac{2}{N} {\cal B}^{ij} A_* K_{ij} + {\cal K}^{ijkl} K_{ij} K_{kl} \right]\,,
\eeq
where we have introduced  the quantity
\beq
A_*\equiv\frac1N (A_0-N^i A_i)\,, 
\eeq
and the extrinsic curvature tensor
\beq
K_{ij}\equiv \frac{1}{2N}\left(\dot\h_{ij}-D_i N_j -D_jN_i\right)\,.
\eeq
The coefficients that appear in (\ref{S_kin}) depend on the six arbitrary functions $\f$ and $\a_I$ of (\ref{action}). They are explicitly given by~\cite{LN1,LN2}
\bea
\label{A}
\A & = & \aa+\ab-(\ac+\ad)\An^2+ \ae \An^4\,, \qquad
{\cal B}^{ij}  = \ba \h^{ij} + \bb \tA^i \tA^j \,,
\\
{\cal K}^{ij,kl} & = &  \ka \h^{i(k}\h^{l)j} +\kb\, \h^{ij} \h^{kl}+  \frac12 \kc\left(\tA^i\tA^j \h^{kl}+\tA^k \tA^l \h^{ij}\right)
    \cr
  &&
  +\frac 12 \kd  \left(\tA^i\tA^{(k}\h^{l) j}+\tA^j\tA^{(k} \h^{l)i}\right) 
  +\ke \tA^i \tA^j\tA^k\tA^l\,,
\eea
 with 
 \begin{eqnarray}
&& \ba=\frac{\An}{2}(2\ab - \ac \An^2 + 4\fX) \,, \quad \bb=\frac{\An}{2} (2\ae \An^2 - \ac - 2\ad)  \,,
\\
&&\ka=\aa \An^2 + f\,, \, \kb=  \ab \An^2-f\,, \,    \kc=- \ac\An^2+4\fX\,, \,  \kd=- 2\aa\,,\,  \ke=\ae\An^2-\ad\,.
\label{k}
 \end{eqnarray}
 The three-dimensional vector $\tA^i$ is defined by $\tA_i\equiv A_i$ and $\tA^i\equiv h^{ij}\tA_j$.
 
  By choosing an appropriate basis of  the six-dimensional vector space of symmetric $3\times 3$ matrices, where the $K_{ij}$ take their values,  the kinetic matrix associated with (\ref{S_kin}) can  be written as a $7\times 7$ block diagonal symmetric matrix of  the form~\cite{LN2}
 \beq
 \left(
\begin{array}{cc}
{\cal M} & \bf{0}\\
\bf{0} & {\cal D}
\end{array}
\right)\,,
\eeq
with the $3\times 3$ matrix 
\bea\label{M}
{\cal M}\equiv \left(
\begin{array}{ccc}
{\cal A} & \frac{1}{2}(\beta_1 + \tA^2 \beta_2) & \frac{1}{\sqrt{2}} \beta_1 \\
\frac{1}{2}(\beta_1 + \tA^2 \beta_2) & \ka + \kb + \tA^2(\kc + \kd) + (\tA^2)^2 \ke  &\sqrt{2}(\kb + \frac{1}{2} \tA^2 \kc) \\
\frac{1}{\sqrt{2}} \beta_1 & \sqrt{2}(\kb + \frac{1}{2} \tA^2 \kc) & \ka + 2 \kb
\end{array}
\right),
\eea
and the diagonal matrix 
\beq
{\cal D}={\rm Diag}\left[\ka, \ka, \ka+\frac12\tA^2\kd ,\ka+\frac12\tA^2\kd \right]\,.
\eeq
The coefficients in the first line (or first row)  of ${\cal M}$ describe the kinetic terms associated with  the scalar field related variable $\An$, including its mixing with the metric sector. As for the metric sector alone, it  is described by the right lower $2\times 2$ submatrix of ${\cal M}$, which we will call ${\cal M}_K$,  together with ${\cal D}$. As our goal is to eliminate the extra degree of freedom due to the higher derivatives of the scalar field, we are looking for a degeneracy of the kinetic matrix that arises from the scalar sector. As a consequence, we will be interested in theories such that ${\cal M}$ is degenerate, while ${\cal M}_K$ and ${\cal D}$ remain nondegenerate in order to preserve the usual tensor structure of gravity.

Requiring the determinant of the matrix ${\cal M}$ to vanish\footnote{Note that we have not used the same matrix in \cite{LN1} but another, non symmetric, matrix constructed  by solving for null eigenvectors of the kinetic matrix. The two methods are obviously equivalent.}
 yields an expression of the form
\beq
\label{determinant}
D_0(X)+D_1(X) \An^2+D_2(X) \An^4=0\,,
\eeq
where we have substituted the expressions (\ref{A})-({\ref{k}) into (\ref{M}) and replaced all $\tA^2$ by $X+\An^2$. 
The functions $D_0$, $D_1$ and $D_2$ depend on the six arbitrary functions $\tf$ and $\a_I$ of the initial Lagrangian:
\beq
\label{D0}
D_0(X)\equiv -4 (\aa+\ab) \left[X \f (2\aa+X\ad+4\f_X)-2\f^2-8X^2\f_X^2\right]\,,
\eeq
\begin{eqnarray}
D_1(X)&\equiv& 4\left[X^2\aa (\aa+3\ab)-2\f^2-4X\f \ab\right]\ad +4 X^2\f(\aa+\ab)\ae 
\cr
&&
+8X\aa^3-4(\f+4X\f_X-6X\ab)\aa^2 -16(\f+5X \f_X)\aa \ab+4X(3\f-4X \f_X) \aa\ac 
\cr
&&
-X^2\f \ac^2 +32 \f_X(f+2X f_X) \ab-16\f \f_X \aa-8\f (\f-X\f_X)\ac+48\f \f_X^2 \,,
\end{eqnarray}
\begin{eqnarray}
D_2(X)&\equiv& 4\left[ 2\f^2+4X\f \ab-X^2\aa(\aa+3\ab)\right]\ae  + 4\aa^3+4(2\ab-X\ac-4\f_X)\aa^2+3X^2 \aa\ac^2
\cr
&&
-4X\f \ac^2+8 (\f+X\f_X)\aa\ac -32 \f_X \aa\ab+16\f_X^2\aa
+32\f_X^2\ab-16\f\f_X\ac\,.
\label{D2}
\end{eqnarray}
Since the determinant must vanish for any value of $\An$, we deduce that degenerate theories are characterized by the three conditions
\beq
D_0(X)=0, \qquad D_1(X)=0, \qquad D_2(X)=0\,.
\eeq
By solving these three conditions, one can determine and classify all  DHOST theories, as discussed in \cite{LN1}.

\subsection{Degenerate theories}
\label{Section_classification}
The  condition $D_0(X)=0$ is the simplest of all three and allows to distinguish several  classes of theories. Indeed,  $D_0$ can vanish either if 
 $\aa+\ab=0$, which defines our first class of solutions, or  if the  term between brackets in (\ref{D0}) vanishes, which defines our second class, as well as our third class corresponding to the special case where $f=0$.

\subsubsection{Class I  ($\aa+\ab=0$)}
\label{subsectionA}  
This class is characterized by the property
\beq
\aa=-\ab\,.
\eeq
One can then  use the conditions  $D_1(X)=0$  and  $D_2(X)=0$ to express, respectively, $\ad$  and $\ae$ in terms of $\ab$ and $\ac$, provided $f+X\ab\neq 0$. This defines the subclass Ia, characterized by
\begin{eqnarray}
\label{a4_A}
\ad&=&\frac{1}{8(f+X\ab)^2}\left[16 X \ab^3+4 (3\f+16 X\f_X)\ab^2
+(16X^2 \f_X-12X\f) \ac\ab-X^2\f \ac^2
\right.
\cr
&&\qquad\qquad \qquad 
\left.
+16 \f_X(3\f+4X\f_X)\ab+8\f (X\f_X-\f)\ac+48\f \f_X^2\right]
\end{eqnarray}
and
\beq
\label{a5_A}
\ae=\frac{\left(4\f_X+2\ab+X\ac\right)\left(-2\ab^2+3X\ab\ac-4\f_X \ab+4\f \ac\right)}{8(f+X\ab)^2}\,.
\eeq
Degenerate theories in class Ia thus depend on three arbitrary functions $\ab$, $\ac$ and $f$. 

In the special case $f+X\ab= 0$,  we find another subclass of solutions characterized by
\beq
\aa=-\ab=\frac{\f}{X}\,, \qquad \ac=\frac{2}{X^2}\left(f-2X f_X\right),\qquad ({\rm Class\  Ib}),
\eeq
where $\f$, $\ad$ and $\ae$ are arbitrary functions. In the following, we will not explore this class much further because  the metric  sector is degenerate. Indeed,   the last two eigenvalues of ${\cal D}$, which are equal  to $f-\aa X$,  vanish in this case.

\subsubsection{Class II}
The condition $D_0(X)=0$  can also be  satisfied if 
\beq
\label{cond3}
X \f (2\aa+X\ad+4\f_X)-2\f^2-8X^2\f_X^2=0 \,.
\eeq
We can then proceed as previously by solving $D_1(X)=0$ and $D_2(X)=0$ to express $\ad$ and  $\ae$ in terms of the three other functions. Substituting the obtained expression for $\ad$ into the condition (\ref{cond3}),  one finally gets
\beq
 (X\aa -f)\left[(4f^2 + X f (8\ab  + 2 \aa  + X\ac-4\f_X)-4X^2 f_X(\aa+3\ab)\right]=0\,.
\eeq
Assuming that $f- X\aa \neq 0$, this   leads to the expressions
\begin{eqnarray}
   \ac &=&\frac{1}{X^2 f}\left[-4 f (f - X f_X )
   -2 X (f - 2 X f_X )\aa+4 X (-2 f + 3 X f_X )
   \right] \,,
   \\
   \ad & = & \frac{2}{X^2 f}\left[f^2 - 2 f X f_X  + 4 X^2 f_X^2- X f  \aa\right]\,,
   \\
   \ae & = & \frac{2}{f^2 X^3}\left[4 f (f^2 - 3 f X f_X  + 2 X^2 f_X^2)+(3 X f^2  - 8 X^2 f f_X  + 6 X^3  f_X^2)\aa
   \right. 
   \nonumber
   \\
&& \left. \qquad \qquad   +2 X (2 f - 3X  f_X )^2\ab\right]\,,
   \label{classII}
\end{eqnarray}
while  $\tf$, $\aa$ and $\ab$ are arbitrary. This describes our class IIa, characterized by three arbitrary functions.

The case $f=X\aa$ defines another class, similar to class Ib, which we will call class IIb, described by
\begin{eqnarray}
 \aa&=&\frac{f}{X}\,,\qquad \ad=4 f_X\left(2\frac{f_X}{f}-\frac{1}{X}\right)\,,
 \\
    \ae&=&\frac{1}{4 X^{3} f ( f + X \ab ) }\left[
    8 X ( 4 X f_{X} f - f^{2} - 4 X^{2} f^{2}_{X} ) \ab + X f ( 8 X^{2} f_{X} + X^{3} \ac - 4 f) \ac 
     \right. 
   \nonumber
   \\
   && \left. \qquad \qquad \qquad \qquad 
    + 4 ( X f_{X} f^{2} - 2 X^{3} f^{3}_{X} + 2 X^{2} f^{2}_{X} f - f^{3} )
    \right]\,,
   \label{classIIa}
\end{eqnarray}
where $\ab$ and $\ac$ are arbitrary functions. Like class Ib,  the metric sector is degenerate for these theories and we will not consider them further  in the following.

\subsubsection{Class III ($f=0$)}
Finally, we devote a special class to the case $f=0$, which also leads automatically to $D_0=0$. Using $D_1=0$ and $D_2=0$ to determine $\ad$ and $\ae$, one gets 
\beq
\ad = -\frac{2}{X}\aa\,, \qquad  \ae=\frac{4 \aa^2+8 \aa
   \ab-4 \aa \ac X+3 \ac^2 X^2}{4 X^2 (\aa+3
   \ab)}  \qquad {\rm (class\  IIIa)}\,,
 \eeq
 while $\aa$, $\ab$ and $\ac$ are arbitrary, provided $\aa+3
   \ab\neq 0$. This  defines our class IIIa. 
   Note that the intersection of IIIa with the class Ia is described by
  \beq
  \ab=-\aa\,, \quad \ad = -\frac{2}{X}\aa\,, \qquad \ae=\frac{(2\aa-X\ac)(2\aa+3 X\ac)}{8 X^2 \aa}\,, \qquad ( {\rm   IIIa} \cap {\rm  Ia} )\,,
  \eeq
  which depends on two arbitrary functions, $\aa$ and $\ac$, and  includes the Lagrangian $L_4^{\rm bh}$ (for which $\aa/X=\ac/2=F_4$). 
   
The case $\aa+3\ab= 0$ yields another subclass,  
\beq
f=0\,,\qquad \aa=\frac32 X\ac,\qquad \ab=-\frac{X}{2}\ac  \qquad {\rm (class\  IIIb)}\,,
\eeq
which in general leads to a degenerate metric sector. 
Another special case  corresponds to the class 
\beq
f=0\,,\qquad \aa=0\,,   \qquad {\rm (class\  IIIc)}
\eeq
which depends on four arbitrary functions. Since $f-\aa X=0$, this class  is also degenerate in the metric sector.

\subsubsection{Degeneracy of the scalar sector alone}
Among all the degenerate theories that we have listed above, it is not difficult to  identify the theories that remain degenerate even when the metric becomes  nondynamical, as noted in \cite{LN1}. In this limit, only the kinetic term for $\An$ is relevant and the degeneracy of the scalar sector alone thus  requires  $\A=0$, which imposes simultaneously the three constraints 
\beq
\label{deg_scalar}
\aa+\ab=0\,,\qquad \ac+\ad=0\,, \qquad \ae=0\,.
\eeq
The first condition implies that the theories satisfying these conditions belong to class I. Ignoring class Ib, whose metric sector is degenerate, we turn to  class Ia.  For theories satisfying (\ref{deg_scalar}),  the functions $\ab$ and $\ac$ are no longer independent, but related by
\beq
\label{cond2}
4\f_X+2\ab+X\ac=0\,.
\eeq
This means that the condition $\A=0$ restricts the degenerate theories to a subclass that depends on two arbitrary functions only. 
It is easy to see that this family of theories in fact coincides with the sum of $L_4^{\rm H}$ and  $L_4^{\rm bH}$, upon using  the identification
\beq
\f=G_4\,,\qquad  \aa=-\ab=2 G_{4X}+ X F_4\,, \qquad \ac=-\ad=2 F_4\,.
\eeq
This implies that the quartic Lagrangian $L_4=L_4^{\rm H}+L_4^{\rm bH}$ represents  the most general theories  that are  degenerate when the metric is nondynamical (with a nondegenerate metric sector).

\section{Disformal transformations}
\label{section_disformal}
We now study the effect of conformal-disformal transformations, or generalized disformal transformations, introduced in \cite{Bekenstein:1992pj}, in which the ``disformed''  metric $\tg_{\mu\nu}$ is expressed in terms of $g_{\mu\nu}$ and $\phi$ as
\beq
\label{disformal}
\tg_{\mu\nu}=\C(X, \phi) g_{\mu\nu}+\D(X, \phi) \, \phi_\mu\, \phi_\nu\,.
\eeq
Via this transformation,  any action $\tilde S$ given as a functional  of  $\tg_{\mu\nu}$ and $\phi$ induces a new   action $S$  for  $g_{\mu\nu}$ and $\phi$, when one substitutes the above expression for  $\tg_{\mu\nu}$ in $\tilde S$: 
 \beq
S[\phi, g_{\mu\nu}]\equiv\tilde S\left[\phi, \tg_{\mu\nu}=\C \,g_{\mu\nu}+\D \, \phi_\mu\phi_\nu\right]\,.
\eeq
We will  say that the actions $S$ and $\tilde{S}$ are related by the disformal transformation (\ref{disformal}).

Starting from an action $\tilde S$ of the form (\ref{action}), 
\beq
\tilde S=\tilde S_g+\tilde S_\phi=\int d^4x\sqrt{-\tg}\left[\, \tf\, \tilde R+\ta_I \tilde L_I^\phi\right]\,,
\eeq
we show below that the action $S$, related to $\tilde S$ via a disformal transformation, is also of the form (\ref{action}), up to terms of the form (\ref{S_other}), and we compute explicitly the relations between the functions that appear in the two Lagrangians.
Interestingly, if the disformal transformation is invertible, in the sense that the metric $g_{\mu\nu}$ can be expressed in terms of $\tg_{\mu\nu}$, then the number of degrees of freedom associated with $S$ and $\tilde S$ should be the same. One thus expects that the disformal transformations of all the degenerate theories described in the previous section are also degenerate. 
We will also discuss the special case where the transformation is non invertible in subsection \ref{subsection_mimetic}.

\subsection{Relations between the two metrics and their covariant derivatives}
In order to write explicitly the above action in terms of $g_{\mu\nu}$ and $\phi$, we will need the expression of the inverse metric 
\beq
\tg^{\mu\nu}=\C^{-1}\left(g^{\mu\nu}-\frac{\D}{\C+\D X}\nabla^\mu\phi\nabla^\nu\phi\right)\,.
\eeq
Contracting this relation with $\phi_\mu\, \phi_\nu$ gives  $\tilde X$ as a function of $X$:
\beq
\label{X_tilde}
\tilde X=\frac{X}{\C+\D X}\,.
\eeq
It is also useful to introduce the ratio
\beq
{\cal J}_g\equiv \frac{\sqrt{-\tg}}{\sqrt{-g}}=\C^{3/2}\, \sqrt{\C+\D X} \,.
\eeq

The difference between the two covariant derivatives $\tilde\nabla$ and $\nabla$, associated respectively to the two metrics $\tg_{\mu\nu}$ and $g_{\mu\nu}$, is fully characterized by the  difference of their respective Christoffel symbols, 
\beq
C_{\mu\nu}^\lambda\equiv\tilde\Gamma_{\mu\nu}^\lambda-\Gamma_{\mu\nu}^\lambda\,, 
\eeq
which defines a tensor. In particular, the relation between the respective second order covariant derivatives of $\phi$ reads
\beq
\tnabla_\mu\tnabla_\nu\phi=\nabla_\mu\nabla_\nu\phi-C_{\mu\nu}^\lambda\phi_\lambda\,.
\eeq

 The explicit expression for $C_{\mu\nu}^\lambda$  is given by
\begin{eqnarray}
\label{C}
C_{\mu\nu}^\lambda&=&\frac{\C_X}{\C}\left[2\d^\lambda_{(\mu}\phi_{\nu)\sigma}\phi^\sigma-\phi^{\lambda\sigma}\phi_\sigma g_{\mu\nu}+\frac{\D}{\C+\D X}\left(-2\phi^\lambda\phi_{(\mu}\phi_{\nu)\sigma}\phi^\sigma+\phi^\lambda \phi^\rho\phi_{\rho\sigma}\phi^\sigma  g_{\mu\nu}\right)\right]
\nonumber
\\
&&+\D_X\left[-\frac{1}{\C}\phi_\mu\phi_\nu\phi^{\lambda\sigma}\phi_\sigma
+\frac{1}{\C+\D X}\left(2\phi^\lambda\phi_{(\mu}\phi_{\nu)\sigma}\phi^\sigma+\frac{\D}{\C}\phi^\rho\phi_{\rho\sigma}\phi^\sigma \phi^\lambda\phi_\mu\phi_\nu\right)\right]
+\frac{\D}{\C+\D X}\phi^\lambda\phi_{\mu\nu}
\nonumber
\\
&&
+\frac{A_\phi}{2 A }\left[ \d^\lambda_\mu \phi_\nu +\d^\lambda_\nu\phi_\mu- \frac{1}{A+B X}(A \phi^\lambda g_{\mu\nu}+ 2B \phi^\lambda \phi_\mu\phi_\nu)\right]+\frac{B_\phi}{2(A+B X)}\phi^\lambda \phi_\mu \phi_\nu\,.
\end{eqnarray}
The last line does not depend on second derivatives of $\phi$. As we will see, this implies that the terms in $A_\phi$ and $B_\phi$ appear only in the transformed action as terms of the form (\ref{S_other}), which we will not compute explicitly.

\subsection{Curvature term}
Let us first concentrate, in $\tilde S$,  on the term depending on the Ricci scalar of $\tg_{\mu\nu}$. Following  the derivation presented in \cite{Zumalacarregui:2013pma},
the Ricci scalar $\tilde R$   can be written in terms of the tensor $C_{\mu\nu}^\lambda$ and of the metric $g_{\mu\nu}$, according to the expression
\beq
\tilde R\equiv \tg^{\mu\nu}\tilde R_{\mu\nu}
=\C^{-1}\left(g^{\mu\nu}-\frac{\D}{\C+\D X}\phi^\mu\phi^\nu\right)
\left(R_{\mu\nu}+C_{\mu\rho}^\sigma C_{\nu\sigma}^\rho-C_{\mu\nu}^\rho C_{\rho\sigma}^\sigma\right)+\tnabla_\rho \xi^\rho\,,
\eeq
with 
\beq
\xi^\rho\equiv \tg^{\mu\nu}C_{\mu\nu}^\rho -\tg^{\rho\mu} \, C_{\mu\nu}^\nu\,.
\eeq
All the terms quadratic in $C_{\mu\nu}^\lambda$ can be rewritten in terms of the elementary Lagrangians 
$L^\phi_I$. One finds
\beq
\label{CC}
\C^{-1}\left(g^{\mu\nu}-\frac{\D}{\C+\D X}\phi^\mu\phi^\nu\right)\left(C_{\mu\rho}^\sigma C_{\nu\sigma}^\rho-C_{\mu\nu}^\rho C_{\rho\sigma}^\sigma\right)=\sum_I \r_I L_I^\phi +(\dots)\,,
\eeq
with 
\bea
\r_1 &=& \r_2=0\,, \qquad \r_3=-\frac{B \left(B X A_X+A \left(2
   A_X+X B_X+B\right)\right)}{A^2
   (A+B X)^2}\,,
   \\
\r_4&=& \frac{(6 A^2 + 8 A B X + 2 B^2 X^2) A_X^2+4 A X (A + B X) A_X B_X+A^2 B (B + X B_X)
}{A^3 (A+B
   X)^2}  \,,
 \\
\r_5&=& -\frac{2 A_X \left(B A_X+2 A
   B_X\right)}{A^3 (A+B X)}\,.
\eea
The dots in (\ref{CC}) indicate terms that are  at most linear in $\phi_{\mu\nu}$, i.e. of the form (\ref{S_other}), which we will not write down explicitly. 

The total derivative  $\tnabla_\rho \xi^\rho$ can be ignored if the function $\tilde f$ multiplying the scalar curvature is a constant. Otherwise, one also needs to reexpress this term as a function of $g_{\mu\nu}$ and $\phi$. This can be done after an integration by parts so that one gets  
\beq
\int d^4 x\sqrt{-\tg} \, \tf \tnabla_\mu \xi^\mu=- \int d^4 x\sqrt{-\tg}\, \xi^\mu \nabla_\mu \tf
=- 2\int d^4 x\sqrt{-\tg}\,  \tf_{\tilde X} \, \tX_X \, \xi^\mu\phi_{\mu\nu}\phi^\nu +(\dots)\,,
\eeq
with 
\beq
\tX_X\equiv \frac{\partial\tilde X}{\partial X} =\frac{A}{(A+B X)^2}\,.
\eeq
Since $\xi^\mu$ contains second derivatives of $\phi$, the scalar quantity $\xi^\mu\phi_{\mu\nu}\phi^\nu$ can be decomposed as a combination of the elementary terms $L_I^\phi$. One finds 
\beq
\xi^\mu\phi_{\mu\nu}\phi^\nu=\l_I L_I^\phi +(\dots)\,,
\eeq
with 
\begin{eqnarray}
\l_1 &=& \l_2=0\,,\qquad 
\l_3= \frac{B}{A^2+A B X}\,, 
\\
\l_4&=&-\frac{4 B X A_X+A \left(6 A_X+2 X
   B_X+B\right)}{A^2 (A+B X)}\,, \qquad 
\l_5 = \frac{2 \left(2 B A_X+A B_X\right)}{A^2
   (A+B X)}\,,
\end{eqnarray}   
and the dots stand as usual for the terms at most linear in $\phi_{\mu\nu}$.   
   
Putting everything together, one finds that the scalar curvature term yields
\beq
\int d^4x\sqrt{-\tg}\, \tf\, \tilde R=\int d^4x\sqrt{-g} \, {\cal J}_g\left\{\frac{\tf}{\C}\left[\, R
-\frac{\D}{\C+\D X} R_{\mu\nu}\, \phi^\mu\phi^\nu \right]+\left(\r_I-2 \tX_{X}\tf_{\tilde X} \l_I\right) L_I^\phi\right\}+(\dots)\,,
\eeq
where the term in $R_{\mu\nu}\, \phi^\mu\phi^\nu$ is of the form (\ref{S_Ricci})
with the function
\beq
\label{function_h}
h=-{\cal J}_g\, \frac{\D}{\C(\C+\D X)} \, \tf \,,
\eeq
and the dots correspond to terms of the form (\ref{S_other}).

\subsection{Scalar field terms}
Let us now consider the terms quadratic in second derivatives  of the scalar field. Each of the  five terms in $\tilde S_\phi$ can be decomposed, after  substitution of (\ref{disformal}), into the five terms that appear in the final action $S_\phi$. 

Let us illustrate this with the first term $\tilde L_1^\phi\equiv\tilde\phi_{\mu\nu}\, \tilde\phi^{\mu\nu}$, which can be decomposed as follows:
\begin{eqnarray}
\tilde L_1^\phi&=&\tg^{\mu\rho} \, \tg^{\nu\sigma} \,\tnabla_\mu\!\tnabla_\nu\phi\,  \tnabla_\rho\!\tnabla_\sigma\phi   
\\
&=& \C^{-2}\left(g^{\mu\rho}-\frac{\D}{\C+\D X}\phi^\mu\phi^\rho\right)\left(g^{\nu\sigma}-\frac{\D}{\C+\D X}\phi^\nu\phi^\sigma\right)\left(\phi_{\mu\nu}-C_{\mu\nu}^\lambda\phi_\lambda\right)\left(\phi_{\rho\sigma}-C_{\rho\sigma}^\tau\phi_\tau\right)
\nonumber
\\
&=&
\T_{11}\, L_1^\phi+\T_{13}\, L_3^\phi+\T_{14}\, L_4^\phi+\T_{15}\, L_5^\phi +(\dots)\,,
\end{eqnarray}
where the coefficients are determined explicitly by substituting the expression (\ref{C}) for $C_{\mu\nu}^\lambda$. Note that the term $L_2^\phi$ does not appear in the decomposition.

Proceeding similarly with all the other terms, one finally gets five similar decompositions, which can be summarized by the  expression
\beq
\tilde L^{\phi}_I=\T_{IJ} L^\phi_J +(\dots)\,,
\eeq
where the summation with respect to the index $J$ is implicit.  The nonvanishing coefficients $\T_{IJ}$ are given by
\begin{eqnarray}
\label{T11}
\T_{11}&=& \frac{1}{(A+B X)^2}\,, \quad \T_{13}=\frac{2 A_X}{A (A+B X)^2}\,,\quad  
\T_{14}=\frac{2 \left(X \left(A_X+X
   B_X\right){}^2-A \left(2 \left(A_X+X
   B_X\right)+B\right)\right)}{A (A+B
   X)^3}, 
\nonumber
\\
\T_{15}&=&\frac{1}{A^2
   (A+B X)^4}\left[2 A^3 B_X+A^2 \left(-2 X A_X
   B_X+2 A_X^2+B^2-X^2 B_X^2+4 B X
   B_X\right)+3 B^2 X^2 A_X^2
   \right.
   \cr
   &&\left. \qquad \qquad \qquad \qquad
   -2 A B X
   \left(2 X A_X B_X+A_X \left(B-2
   A_X\right)+X^2 B_X^2\right)\right]
\nonumber
\\
\T_{22}&=&\frac{1}{(A+B X)^2}\,, \quad  
 \T_{23}=-\frac{2 \left(A \left(-2 A_X+X B_X+B\right)-3 B X A_X\right)}{A (A+B X)^3}\,, 
 \nonumber
 \\
   \T_{25}&=& \frac{\left(A \left(-2 A_X+X
   B_X+B\right)-3 B X
   A_X\right){}^2}{A^2 (A+B X)^4}\,,
 \nonumber
 \\
     \T_{33}&=& \frac{A-X \left(A_X+X B_X\right)}{(A+B
   X)^4} \,,\quad
\T_{35}= \-\frac{\left(A \left(-2 A_X+X
   B_X+B\right)-3 B X A_X\right)
   \left(A-X \left(A_X+X
   B_X\right)\right)}{A (A+B X)^5}\,,
 \nonumber
   \\
    \T_{44}&=& \frac{\left(A-X \left(A_X+X
   B_X\right)\right){}^2}{A (A+B X)^4}\,,\quad  
   \T_{45}=-\frac{B \left(A-X \left(A_X+X B_X\right)\right){}^2}{A (A+B X)^5}\,, 
   \nonumber
   \\
   \T_{55}&=&\frac{\left(A-X \left(A_X+X B_X\right)\right){}^2}{(A+B X)^6}\,.
\end{eqnarray}
It can be noticed that these coefficients form a triangular matrix. 

\subsection{Transformation of the total action}
Collecting all the results obtained above, one can now write the functions that appear in the action $S$ in terms of the functions $\tf$ and $\ta_I$ of $\tilde S$. 
We find
\begin{eqnarray}
\label{f_gen}
f&=& {\cal J}_g\, \C^{-1} \tf\,,
\\
\aa&=&-h +{\cal J}_g \,\T_{11} \,\ta_1\,,
\label{a1_gen}
\\
\ab&=&h +{\cal J}_g \,\T_{22} \,\ta_2\,,
\\
\ac&=& 2 h_X+{\cal J}_g \left[ \tf \r_3 
-2\tX_X \tf_{\tX}\l_3+\T_{13} \,\ta_1+\T_{23} \,\ta_2+\T_{33} \,\ta_3
\right]\,,
\\
\ad&=& - 2 h_X+{\cal J}_g \left[ \tf \r_4 
-2\tX_X \tf_{\tX}\l_4+\T_{14} \,\ta_1+\T_{44} \,\ta_4
\right]\,,
\\
\ae&=&{\cal J}_g \left[ \tf \r_5
-2\tX_X \tf_{\tX}\l_5+\T_{15} \,\ta_1+\T_{25} \,\ta_2+\T_{35} \,\ta_5+\T_{45} \,\ta_5+\T_{55} \,\ta_5
\right]\,.
\label{a5_gen}
\end{eqnarray}
By substituting all the formulas given in the previous subsections, one obtains the explicit expressions of $\f$ and $\a_I$ in terms of $\tf$, $\ta_I$, $\C$ and $\D$. One can verify that the $\f$ and $\a_I$ satisfy the degeneracy conditions (\ref{D0})-(\ref{D2}). In fact, it turns out that this is a very efficient way to check the expressions for $\f$ and $\a_I$. A first conclusion is thus that all quadratic DHOST theories transform into quadratic DHOST theories.

For a more detailed analyis, the relations (\ref{f_gen})-(\ref{a5_gen})  in their abridged form are useful to see how the main families of DHOST theories  transform. First of all, let us note that if $\tf=0$ then necessarily $f=0$. Therefore, the transformed version  of  theories in class III remains in class III. 
As a consequence of $\T_{11}=\T_{22}$, we also   find the relation
\beq
\aa+\ab={\cal J}_g \,\T_{11} \, (\ta_1+\ta_2)\,,
\eeq
which shows that the property $\aa+\ab=0$ (or $\aa+\ab\neq 0$) is unchanged by disformal transformations. 
This implies  that class I, characterized by $\aa+\ab=0$, is  stable under disformal transformations. Therefore all the three main classes are stable. 
We study more precisely  the impact of disformal transformations  in the next two sections.

\section{Disformal transformations in Class Ia}
\label{section_classI}
Disformal transformations  for theories in  class Ia have been partially studied in several previous works. In particular, it has been shown that Horndeski theories are stable under $X$-independent disformal transformations~\cite{Bettoni:2013diz}. The first example of theory beyond Hordenski, i.e. with higher order equations of motion, was exhibited in \cite{Zumalacarregui:2013pma} by considering the general disformal transformation of the Einstein-Hilbert action, which is also in class Ia. It was also shown  in \cite{Gleyzes:2014qga} that the extended quadratic (quintic) Lagrangian proposed in \cite{Gleyzes:2014dya}
  can be generated from the quadratic Horndeski Lagrangian via a purely disformal transformation with $\C=1$. All these examples are particular cases of disformal transformations within class Ia.
 
If we now consider class Ia theories such that $\tf\neq 0$, which depend on three arbitrary functions, it is natural to expect that generic theories  can be ``generated'' from the subset of (quadratic) Horndeski theories, characterized by a single arbitrary function $\tf$, via general disformal transformations, which depend on two arbitrary functions.  We can check that this is indeed the case\footnote{The same calculation has been performed independently in the recent paper \cite{Crisostomi:2016czh}.}, by starting from the quartic Horndeski Lagrangian expressed in terms of the metric $\tg_{\mu\nu}$ and of the scalar field $\phi$,
\beq
\label{S_tilde}
\tilde S[\phi, \tg_{\mu\nu}]=\int d^4 x\sqrt{-\tg} \left\{\tf (\tilde X,\phi) \tR - 2 \tf_{,\tilde X}(\tilde X, \phi)\left[ (\tnabla^\mu \tnabla_\mu\phi)^2 - \tnabla_\mu\tnabla_\nu\phi\, \tnabla^\mu \tnabla^\nu \phi\right]
\right\}\,.
\eeq
Substituting (\ref{disformal}), we obtain an action $S$  for $g_{\mu\nu}$ and $\phi$, which is characterized by the functions
\beq
\label{f_gen_I}
f=A^{1/2}\,  \sqrt{A+B X}\, \tf\,,
\eeq
and
\begin{eqnarray}
\aa&=&-\ab = -\frac{2 A^{3/2} }{(A+B X)^{3/2}} \, [ ( B + XB^{2} )\tf + 2 \tf_{\tX} \, ]
\label{a1_gen_I}
\\
\ac&=& -\frac{2 \left(B A_X+A
   B_X\right)}{A^{1/2} (A + BX)^{1/2}}\tf 
   +\frac{4 ( X B_X- A_X) A^{1/2}}{(A+B X)^{3/2}}
\\
\ad&=& \frac{2 \left(A^2 B_X+A A_X \left(2
   X B_X+B\right)+A_X^2 (3 A+B
   X)\right)}{A^{3/2} ( A + BX )^{1/2}}\tf
   \cr
&&   -\frac{4 
   \left(-A_X \left(A-2 X^2
   B_X\right)+A X B_X+2 X
   A_X^2\right)}{A^{1/2}(A+B X)^{3/2}}\tf_{\tX}
\\
\ae &=&-\frac{2 A_X 
   \left(B A_X+2 A B_X\right)}{A^{3/2}
   (A+B X)^{1/2}}\tf
   +\frac{4 A_X 
   \left(2 X B_X-A_X\right)}{A^{1/2} (A+B
   X)^{3/2}}\tf_{\tX}
    \label{a5_gen_I}
\end{eqnarray}

If one starts from a generic theory in Class I, defined by the functions $\f$ and $\alpha_I$, it is possible to determine two functions $\C$ and $\D$ such that this theory is  disformally related to  Horndeski, as we now show.  According to (\ref{f_gen_I}), the Horndeski function $\tf$ is related to $f$, $\C$ and $\D$ by 
\beq
\label{G2f}
\tf=A^{-1/2} (A + BX)^{-1/2}\,f\,.
\eeq
Substituting this expression  for  $\tf$ into  (\ref{a1_gen_I}) yields
\beq
\aa = -\ab= \frac{2 A f_X- f(2 A_X+X B_X)}{A-XA_X-X^2 B_X},
\eeq
which one can solve to find $B_X$ in terms of $\ab$, $f$ and $A$:
\beq
\label{B2a2}
B_X=\frac{(2f_X+\ab) A- (2 f+X \ab) A_X}{X(f+X\ab)}\,.
\eeq
Substituting (\ref{G2f}) and (\ref{B2a2}) in $\ac$ gives 
\beq
\label{A2a3}
\frac{A_X}{A}=\frac{4f_X+2\ab+X\ac}{4(f+X\ab)}\,.
\eeq
Finally, by substituting successively (\ref{G2f}), (\ref{B2a2}) and (\ref{A2a3}), one can rewrite $\ad$ and $\ae$ 
in terms of $f$, $\ab$ and $\ac$ and check that one recovers exactly the expressions (\ref{a4_A}) and (\ref{a5_A}). This proves that generic  theories in class Ia  are ``generated''  from the Horndeski quadratic Lagrangians  via disformal transformations (\ref{disformal}).

In analogy with the choice between the  ``Jordan frame'' and ``Einstein frame'' for traditional scalar tensor theories, the above construction shows that  theories belonging to class Ia with $f\neq0$ can be defined either in the ``Jordan frame'',
where the metric is minimally coupled to  matter, 
\beq
S_{\rm total}=\int d^4 x\sqrt{-g}\left[\f R +\a_I L_I^\phi\right]+S_m[g_{\mu\nu},\Psi_m]\,,
\eeq
or in the ``Horndeski frame'', where the gravitational part of the action is described by Horndeski,
\beq
\tilde S_{\rm total}=\int d^4 x\sqrt{-\tg} \left\{\tf  \tR - 2 \tf_{,\tilde X}\left[ (\tilde\square\phi)^2 - \tilde\phi_{\mu\nu}\tilde\phi^{\mu\nu}\right]
\right\}+(\dots) +S_m[g_{\mu\nu},\Psi_m]\,.
\eeq
In the ``Horndeski frame'', the matter action is nonminimally coupled, but can be expressed explicitly in terms of the ``Horndeski metric'' by inverting the transformation (\ref{disformal}).

Note that the Einstein-Hilbert Lagrangian, with $\tf$ constant and $\ta_I=0$,  is a particular case of Horndeski. It generates, via disformal transformations, the family characterized by the expressions (\ref{f_gen_I})-(\ref{a5_gen_I}) with  $\tf_{\tX}=0$. If the disformal transformation is invertible, one thus gets a family of scalar-tensor theories which are in fact general relativity in disguise and, as such, 
are doubly degenerate and contain only two tensor modes. Of course, one can always add another term of the form (\ref{S_other}) in the action,  which does not modify the quadratic part of the action (\ref{action}),  in order to break the second degeneracy. One then obtains a degenerate scalar-tensor theory with one scalar mode and two tensor modes. This is precisely how a theory ``beyond Horndeski'' was constructed in \cite{Zumalacarregui:2013pma}.

\section{Disformal transformations in other classes}
\label{section_class_II_III}
We now turn to the other classes of DHOST theories. 

\subsection{Stability of all classes}
We have already pointed out the stability, under disformal transformations, of the sign (including zero)  of $f$ and of  $\a_1+\a_2$, which guarantes the stability of the classes I, II and III separately. We now consider the criteria that distinguish the subclasses within these classes. 

One can first notice the relation 
 \beq
 f-\aa X= {\cal J}_g \left[\frac{1}{A+B X}\tf-\frac{X}{(A+BX)^2} \,\ta_1\right]=
 \frac{ {\cal J}_g }{A+B X}\left(\tf-\tX \,\ta_1\right) \,,
 \eeq
 where we have substituted the expression (\ref{function_h}) for $h$ and the coefficient $\T_{11}$ in (\ref{a1_gen}).
If we start from a theory in Class Ib or in Class IIb, characterized by $\ta_1=\tf/\tX$, the above relation implies that the  disformally transformed theory verifies $\aa=\f/X$ and thus belongs to the same subclass, either Ib or IIb, as the  original theory. Therefore, the classes Ia, Ib, IIa and IIb are separately stable. 

We find the same properties for the subclasses in class III. Indeed, when $f=0$, we have $\aa={\cal J}_g \T_{11}\ta_1$ and $\ab={\cal J}_g \T_{11}\ta_2$. Therefore the signs of $\aa$ and $\aa+3\a_2$ which distinguish the subclasses IIIa, IIIb and IIIc are conserved in a disformal transformation.

In summary, all the classes and subclasses that we have distinguished are separately stable under disformal transformations. In particular, the intersections of two classes or subclasses, when  non empty, are also 
stable. This applies for instance to the intersection of Ia and IIIa, which contains $L_4^{\rm bh}$. 

\subsection{Disformal transformations in Class IIa}
It is straightforward to specialize the general disformal transformations to class IIa. One just needs to impose that the tilted functions $\tf$ and the $\ta_I$ satisfy the properties  (\ref{classII}). 
Since Lagrangians in class IIa depend on three arbitrary functions, one can try to proceed as in class Ia by choosing a particular family that depends on a single arbitrary function and then produce  generic theories by applying a disformal transformation. There is no natural candidate for this one-function family, in contrast with Horndeski.  One could choose for example the family
\begin{eqnarray}
\tf&=&1\,, \quad \ta_1=0\,,\quad 
   \ta_3= -\frac{4}{X^2}(1+2 X\ab)\,,
   \quad
\ta_4= \frac{2}{X^2}\,,\quad 
\ta_5=\frac{8}{X^3}(1+2 X\ab)\,,
   \label{gen_II}
\end{eqnarray}
which depends only on the arbitrary function $\ta_2$.

One  finds that the disformal transformation of this family leads, in particular, to
\beq
\label{f_gen_II}
f=A^{1/2} ( A + BX)^{1/2}
\eeq
and
\begin{eqnarray}
\ab&=& -\frac{A^{1/2}  \left(A (B-\ta_2)+B^2 X\right)}{(A+BX)^{3/2}} 
   \label{a2_gen_II}
\end{eqnarray}
We can solve the first equation to determine $B$ in terms of $A$ and $f$:
\beq
B=\frac{f^2-A^2}{AX}\,.
\eeq
Substituting in (\ref{a2_gen_II}) and solving for $\ta_2$, we get
\beq
\ta_2=\frac{f^2}{A^3 X}\left(f^2-A^2+Xf \ab\right)\,.
\eeq
One can then substitute these relations into the other coefficients obtained by disformal transformation. The coefficient $\ad$, for instance,  is particularly simple:
\beq
\ad= \frac{2}{f X^2}\left(A^2+4X^2 f_X^2-2X f f_X\right).
\eeq
In this way, one can determine the function $A$ in terms of $\ad$ and $f$, and then $B$. Note however that this procedure works only if the above equation can be solved for $A^2$. This means that there will be restrictions on the theories generated by the family (\ref{gen_II}).

\subsection{Disformal transformation in class IIIa}
Although class IIIa contains $L_{4}^{\rm bH}$, which depends on a single function, this Lagrangian cannot generate all theories in IIIa because it belongs to the intersection of IIIa and Ia and any of its transformed Lagrangians will also belong to this intersection. 
This means in particular that $L_{4}^{\rm bH}$ cannot be connected to Horndeski, as already pointed out in \cite{Crisostomi:2016tcp}.

In class IIIa, $h=0$ and therefore the ratio between $\aa$ and $\ab$ remains conserved in a disformal transformation. This implies that one cannot choose a ``seed'' family  with $\ta_1=0$ or $\ta_2=0$ in order  to obtain generic theories of IIIa. Instead, one can try to take, for instance, the family
\beq
\tf=0\,, \quad \ta_1=1\,, \qquad \ta_3=0\,, \quad \ta_4=-\frac{2}{X}\,,\quad \ta_5=\frac{1+2\ta_2}{X^2(1+3\ta_2)}\,,
\eeq
which depends on the arbitrary function $\ta_2$. Applying a disformal transformation on this family, one gets 
\beq
\aa= A^{3/2} (A+BX)^{-3/2},\qquad \ab=A^{3/2} (A+BX)^{-3/2} \ta_2\,,
\eeq
which can be solved to give
\beq
B=A \frac{1-\aa^{2/3}}{X \aa^{2/3}}\,, \qquad \ta_2=A^{-3/2} (A+BX)^{3/2}\ab\,. 
\eeq
Substituting into the expression for $\ac$, one finds 
\beq
\frac{A_X}{A}=\frac{2\aa \ac-4\ac{\aa}_X}{6\aa (\aa+2\ab)}\,.
\eeq
This enables us to determine $A$ and then $B$.

\section{Link with khronometric and mimetic theories}
\label{section_khronon_mimetic}
In this section we show that, aside Horndeski and its extension,   quadratic DHOST theories  also contain, 
as particular cases, other  theories that have already been studied elsewhere. In the first part, we identify  some ``khronometric'' theories  in both class I and class II. In the second part, we discuss ``mimetic'' theories, associated with a non invertible disformal transformation.  

\subsection{Khronometric theories}
\label{subsection_khronon}
Khronometric theories~\cite{Blas:2010hb}  represent a subset of Einstein-aether theories~\cite{Jacobson:2000xp} 
for which the  unit time-like vector $u^\mu$, which defines a special frame, is expressed as the normalized  gradient of a scalar field,
\beq
\label{u_khronon}
u_\mu=\frac{\nabla_\mu\phi}{\sqrt{-X}}\,.
\eeq
Their dynamics is described by the action
\beq
\label{S_khronon}
S=\int d^4x \sqrt{-g}\left(\f R+K^{\mu\nu}_{\ \ \rho\sigma} \nabla_\mu u^\rho\nabla_\nu u^\sigma\right)\,,
\eeq
with 
\beq
\label{K_khronon}
K^{\mu\nu}_{\ \ \rho\sigma}\equiv c_1 g^{\mu\nu}g_{\rho\sigma}+c_2 \d^\mu_\rho\d^\nu_\sigma+c_3\d^\mu_\sigma\d^\nu_\rho+c_4  u^\mu u^\nu g_{\rho\sigma}\,,
\eeq
where the $c_a$ are constant. 
This is the most general  Lagrangian which depends only quadratically on the
vector field $u^\mu$ and it is clearly  invariant under arbitrary scalar field redefinitions $\phi \mapsto \psi(\phi)$.
Since $f$ is constant,  we can set $f=1$ without loss of generality.

Substituting (\ref{u_khronon}) into (\ref{S_khronon})-(\ref{K_khronon}), we find that the action (\ref{S_khronon}) is of the form (\ref{action}), with the functions
\begin{eqnarray}
\label{khronon_functions}
\f&=& 1\,, \quad\aa=-\frac1X (c_1+c_3)\,,\quad  \ab=-\frac1X c_2\,, \quad \ac=\frac{2}{X^2}c_2 \,,
\nonumber
\\
\ad &=& \frac{1}{X^2}(c_1+2c_3+c_4)\,, \quad \ae=-\frac{1}{X^3}(c_2+c_3+c_4)\,.
\end{eqnarray}
One immediately sees that the parameter $c_1$ can be absorbed in $c_3$ and $c_4$ by redefining  $c_3+c_1\rightarrow c_3$ and $c_4-c_1\rightarrow c_4$. In the following, we will thus assume $c_1=0$ without loss of generality (one can always return to the original form by using the inverse redefinitions). 

Among khronometric theories, which represent a subset of higher order theories, one can look for degenerate theories  by examining  the degeneracy conditions (\ref{D0})-(\ref{D2}). 
Substituting (\ref{khronon_functions}), with $c_1=0$,  into the first degeneracy condition yields
 \beq
D_0=\frac{4}{X} \left(c_2+c_3\right)\left(c_4-2\right)=0\,,
  \eeq
  whose solutions yield several families of degenerate khronons.
  
 \subsubsection{Khronons in  Class I} 
 Let us  first consider the case 
  \beq
  c_2+c_3=0\,,
  \eeq
   which corresponds to  class I, since this implies $\aa+\ab=0$.  
  Substituting this condition into $D_1$ and $D_2$, one finds 
  \beq
D_1=X D_2= -\frac{8}{X^2}\left(c_2-1\right)^2 c4\,,
\eeq  
which allows two possibilities. 

The first family,  characterized by 
\beq\label{Khron1}
\aa=-\ab=\frac1X,\quad  \ac=\frac{2}{X^2}, \quad \ad=\frac{1}{X^2}(c_4-2),\quad  \ae=-\frac{c_4}{X^3}\,,
\eeq
with $c_4$ arbitrary, belongs to the class Ib.

The second family, described by 
\beq\label{Khron2}
\aa=-\ab=\frac{c_2}{X},\quad  \ac=\frac{2c_2}{X^2}, \quad \ad=-2\frac{c_2}{X^2},\quad  \ae=0\,,
\eeq
with $c_2$ arbitrary, belongs to class Ia (except the case $c_2=1$ which is also in the previous family).

\subsubsection{Khronometric Class II}
The second possibility to satisfy $D_0=0$  is given by
\beq
c_4=2\,,
\eeq
which leads to
\beq
D_1=X D_2=\frac{8}{X^2}(1+c_3)(3c_2+c_3-2)\,.
\eeq
Once again, we get two families, but now belonging to class II. 

The first family, corresponding to 
$c_3=-1$, is described by
\beq\label{Khron3}
\aa=\frac1X,\quad \ab=-\frac{c_2}{X}\,,\quad  \ac=\frac{2c_2}{X^2}, \quad \ad=0,\quad  \ae=-\frac{c_2+1}{X^3}\,,
\eeq
and depends on the  arbitrary parameter $c_2$. These theories are in IIb, except for the case $c_2=1$, which is in IIa.

The second family, corresponding to $c_3=2-3c_2$, also depends on the single parameter $c_2$:
\beq\label{Khron4}
\aa=\frac{3c_2-2}{X},\quad \ab=-\frac{c_2}{X}\,,\quad  \ac=\frac{2c_2}{X^2}, \quad \ad=\frac{6(1-c_2)}{X^2},\quad  \ae=\frac{2(c_2-2)}{X^3}\,.
\eeq
These Lagrangians belong to IIb, except if $c_2=1$. 

\subsubsection{Disformal transformations}
The set of khronometric theories is stable under the action of disformal transformations of the form 
\bea
\tilde{g}_{\mu\nu} = a \, g_{\mu\nu} + b\, u_\mu u_\nu \,,
\eea
i.e. for $A=a$ and $B=-b/X$,  where $a$ and $b$ are constant. 

 Without loss of generality, one can restrict our analysis to one-parameter transformations such that  $b=a-a^{-1}$, which  preserve $f=1$ in the khronometric action \eqref{S_khronon}. It is straightforward to study the action of disformal transformations of this type on the four degenerate khronometric families identified above. One finds that each theory of the  first family \eqref{Khron1} remains invariant. Each of the three other families is stable, the transformed theory being obtained  by the following modification of the parameter $c_2$:
 \beq
 c_2 \rightarrow (c_2-1)a^2 +1\,.
 \eeq
 In particular, one notes that, in the family   \eqref{Khron2}, any theory with $c_2>0$ can be transformed into general relativity by choosing $a^2=1/(1-c_2)$.

 \subsection{Mimetic theories} 
 \label{subsection_mimetic}
In previous sections, we have  assumed that the disformal transformation \eqref{disformal} is invertible, in the sense  
that one can also  express the metric $g_{\mu\nu}$ in terms of $\tilde{g}_{\mu\nu}$. When we relax
this condition, one obtains the so-called ``mimetic'' theories, in analogy with the first model of this kind,  investigated in \cite{Chamseddine:2013kea},  defined from a non-invertible
disformal transformation of the  Einstein-Hilbert action. 

\subsubsection{Non-invertible disformal transformation and symmetries}
By differentiating the expression
\bea
\tilde{g}_{\mu\nu} = A(\phi,X) g_{\mu\nu} + B(\phi,X) \phi_\mu \phi_\nu\,,
\eea
one obtains
\bea
\label{d_tg}
\delta \tilde{g}_{\mu\nu} = F_{\mu \nu} \, \delta \phi + H_{\mu\nu}^\alpha \, \nabla_\alpha \delta \phi + 
J_{\mu\nu}^{\alpha\beta} \, \delta g_{\alpha\beta} 
\eea
with
\bea
F_{\mu \nu} & = & A_\phi g_{\mu \nu} + B_\phi \phi_\mu \phi_\nu \; ,\\
H_{\mu\nu}^\alpha & = & 2(A_X  g_{\mu\nu} + B_X  \phi_\mu \phi_\nu) \phi^\alpha + 
B (\phi_\nu \delta_\mu^\alpha + \phi_\mu g_\nu^\alpha) \,,
\\
J_{\mu\nu}^{\alpha\beta} &=&
A \delta_{(\mu}^\alpha \delta_{\nu)}^\beta - \phi^\alpha \phi^\beta (A_X g_{\mu\nu} +
B_X \phi_\mu \phi_\nu)\,.
\eea

As discussed in \cite{Zumalacarregui:2013pma}, the  disformal transformation 
is non invertible, i.e. $g_{\mu\nu}$ cannot be determined from $\tilde{g}_{\mu\nu} $, if
 the determinant of  the Jacobian matrix  $J_{\mu\nu}^{\alpha\beta} \equiv \frac{\partial \tilde{g}_{\mu\nu}}{\partial g_{\alpha\beta}} $
vanishes.  This happens when $J_{\mu\nu}^{\alpha\beta} $
admits a  null vector $v_{\alpha\beta}$  such that 
\beq
J_{\mu\nu}^{\alpha\beta} v_{\alpha\beta} = 0 \,. 
\eeq
It is straightforward to check that the combination
\bea\label{nullvectors}
v_{\alpha\beta}=A_X g_{\alpha\beta} + B_X \phi_\alpha\phi_\beta
\eea
is a null vector of the Jacobian matrix, provided the functions $A$ and $B$ verify 
\beq
B_X= \frac{A-XA_X}{X^2}\,.
\eeq
After integration, this yields 
\beq
\label{B_mimetic}
B= - \frac{A}{X} + \mu(\phi)\,,
\eeq
corresponding to the disformal transformation
\beq\label{disformal mimetic}
\tilde{g}_{\mu\nu} = A(\phi, X)\left(g_{\mu\nu} - \frac1X \phi_\mu \phi_\nu\right)+ \mu(\phi)\,  \phi_\mu \phi_\nu\,.
\eeq
Note that if we insert (\ref{B_mimetic}) into (\ref{X_tilde}), one gets $\tX=1/\mu(\phi)$, which shows that $\tX$ does not depend on $X$.

\subsubsection{Mimetic action}
If we  start from an action of the form
\bea\label{generalactionfordisformal}
\tilde{S}[\phi,\tilde{g}_{\mu\nu}] = \int d^4x \; \sqrt{-\tg} \left( \tf(\phi) \tilde{R} + 
\alpha_I(\phi) \tilde{L}_I^\phi \right)\,,
\eea
and substitute (\ref{disformal mimetic}), we obtain a new action $S$, given  as a functional of $g_{\mu\nu}$ and $\phi$. 
This leads to a  subclass of our DHOST theories with  particular properties. This procedure has been used  in  \cite{Chamseddine:2013kea} for the Einstein-Hilbert action, i.e. $\tf=1$ and $\ta_I=0$, with the disformatl transformation characterized  by $A=X$ and $B=0$, to introduce the model of mimetic dark matter.  It has been extended in \cite{Deruelle:2014zza}  to a general non-invertible transformation, with (\ref{B_mimetic}). 
In contrast with the generic case where the disformal transformation is invertible,  the number of degrees of freedom is not necessarily the same for $\tilde{S}$ and $S$. In particular, if $\tilde{S}$ is the Einstein-Hilbert action, with only two degrees of freedom, one ends up with three degrees of freedom for $S$, as discussed in  
\cite{Chamseddine:2013kea} and \cite{Deruelle:2014zza}.

Interestingly, the mimetic action $S$  is invariant under the local symmetry
\bea\label{mimetic symmetry}
\delta \phi = 0 \qquad \text{and} \qquad 
\delta g_{\mu\nu} =\varepsilon\,  v_{\mu\nu}= \varepsilon (A_X g_{\mu\nu} + B_X \phi_\mu\phi_\nu)\,,
\eea
where $\varepsilon$  is an infinitesimal space-time function. This symmetry follows immediately from (\ref{d_tg}), together with the property that $v_{\mu\nu}$ is a null eigenvector of the Jacobian matrix.

In the Hamiltonian framework, such a symmetry implies the existence of  an extra first class 
constraint in addition to the usual Hamiltonian and momentum constraints associated with diffeomorphism invariance. This is in contrast with  the standard quadratic DHOST theories, for which the extra constraints are 
generically second class, as  shown in \cite{LN2}. One of these second class constraints is necessary to eliminate the Ostrogradski ghost and we  thus  expect that, even though mimetic theories contain three degrees of freedom, the Ostrogradski ghost is still present.

This is indeed the case for the simplest model of mimetic gravity, obtained from Einstein-Hilbert  with $A=X$ and $B=0$. 
In that case,  the symmetry \eqref{mimetic symmetry} reduces to an invariance under conformal transformations of $g_{\mu\nu}$.
In the Hamiltonian description, this symmetry is necessarily associated to a first class constraint. 
Following the analysis of \cite{LN2}, and introducing the conjugate momenta $\pi^{ij}$ and $p_*$ of  $\h_{ij}$ and $\An$, respectively, one can show that the primary constraint reduces to
\bea
\Psi \equiv \gamma_{ij} \pi^{ij} - \frac{1}{2} p_*\,,
\eea
which is indeed the generator of infinitesimal conformal transformations. As a consequence, the primary constraint is first class and it Poisson commutes with the Hamiltonian and momentum constraints. Hence, there is no  secondary  constraint that eliminates the Ostrogradski ghost. This has already been noticed   in  \cite{Chaichian:2014qba} and  we expect this to remain true for any mimetic-like theory.

\section{Conclusions}
The degeneracy, in the generalized sense introduced in \cite{LN1},  of scalar-tensor Lagrangians  is  a powerful tool to classify viable alternative theories of gravity. In the present work we have revisited all the quadratic DHOST theories identified in \cite{LN1} and studied how they transform under generalized disformal transformations. In order to do so,  we have  obtained the general transformation laws of the six arbitrary functions that appear in the Lagrangian of these theories. 
This shows that  any quadratic DHOST theory is transformed, via disformal transformation, into another quadratic  DHOST theory, up to terms that are at most linear in $\phi_{\mu\nu}$ (which  do not affect the degeneracy of the theory). 
Moreover, we have found that the three  main classes of quadratic DHOST theories, as well as the two or three subclasses within each, are all stable under disformal transformations. One of these subclasses (class Ia) contains the theories  ``generated''  from the (quadratic) Horndeski Lagrangian via disformal transformations, .

Two  disformally related theories describe  distinct physics if matter is assumed to be minimally coupled for both theories. Conversely, a given scalar tensor theory can be described by  different disformally related Lagrangians, provided the coupling to matter is modified accordingly.  In this sense, the situation is very similar to traditional scalar-tensor theories for which one can use the Jordan frame, in which matter is minimally coupled to the metric, or the Einstein frame, where the gravitational dynamics is described by the usual Einstein-Hilbert Lagrangian but at the price of a non-minimal coupling of matter to the metric. 
Similarly, for the quadratic DHOST theories in class Ia, one can work either  in the  Jordan frame, where matter is minimally coupled  but  the equations of motion are in general higher order, or in the ``Horndeski'' frame where the equations of motion are second order but with a non-minimal coupling of matter to the metric. 

Apart Horndeski and its extension, we have also recognized other known theories among the quadratic DHOST theories. Khronometric theories, which are a sub-class of Einstein-aether theories where the unit vector is proportional to the gradient of a scalar field, lead to higher order scalar-tensor theories when covariantized. These covariantized theories are in general not degenerate but we have found that a subset of them are indeed degenerate. 
Our theories also encompass mimetic gravity, which is obtained from the Einstein-Hilbert Lagrangian via a specific, non invertible, disformal transformation. 

Quadratic DHOST theories also contain theories that cannot be related, up to disformal disformations, to known theories, at least to our knowledge,  and thus seem to represent genuinely new scalar-tensor theories,  independently of their specific coupling to matter. It would be interesting to study the cosmology of these new theories by using, for instance,  the general formalism developed in \cite{Gleyzes:2013ooa,Gleyzes:2014rba}.


\vskip 1cm

Note added: the paper \cite{Crisostomi:2016czh}, which appeared on arXiv during the preparation of this manuscript, also studies the theories introduced in   \cite{LN1} and has some partial overlap with the present work, in particular concerning the disformal transformations in class Ia.

\end{document}